\RequirePackage[l2tabu, orthodox]{nag}	
\documentclass[a4paper, 12pt]{article} 

\usepackage[a4paper]{geometry}
\usepackage{fullpage}					
\usepackage{fancyhdr}
\usepackage{acronym}
\normalfont
\usepackage[T1]{fontenc}
\usepackage[utf8]{inputenc}				
\usepackage[swedish,english]{babel}     
\usepackage{xcolor}						

\usepackage{amsmath}
\usepackage{amsfonts}					
\usepackage{amssymb}                    
\usepackage{amsthm}                     
\usepackage{siunitx}					
\usepackage{cancel}						

\graphicspath{{images/}}				
\usepackage{tikz}						
\usepackage{graphicx}					
\usepackage{epstopdf}					
\usepackage{pgfplots}					
\usepackage{multirow}					
\usepackage{multicol}					
\usepackage{booktabs}					

\usepackage[round,authoryear]{natbib} 					
\usepackage[colorlinks=true,linkcolor=blue, pdfborder={0 0 0}]{hyperref}	
\usepackage{cleveref}

\makeatletter 														

\DeclareSIUnit\parsec{pc}
\DeclareSIUnit\lightyears{ly}

\newcommand{\bs}[1]{\ensuremath{\mbox{\boldsymbol$ #1 $}}} 			
\renewcommand{\d}[2]{\frac{d #1}{d #2}} 							

\let\jnl@style=\rm
\def\ref@jnl#1{{\jnl@style#1}}

\def\aj{\ref@jnl{AJ}}                   
\def\actaa{\ref@jnl{Acta Astron.}}      
\def\araa{\ref@jnl{ARA\&A}}             
\def\apj{\ref@jnl{ApJ}}                 
\def\apjl{\ref@jnl{ApJ}}                
\def\apjs{\ref@jnl{ApJS}}               
\def\ao{\ref@jnl{Appl.~Opt.}}           
\def\apss{\ref@jnl{Ap\&SS}}             
\def\aap{\ref@jnl{A\&A}}                
\def\aapr{\ref@jnl{A\&A~Rev.}}          
\def\aaps{\ref@jnl{A\&AS}}              
\def\azh{\ref@jnl{AZh}}                 
\def\baas{\ref@jnl{BAAS}}               
\def\bac{\ref@jnl{Bull. astr. Inst. Czechosl.}}
\def\caa{\ref@jnl{Chinese Astron. Astrophys.}}
\def\cjaa{\ref@jnl{Chinese J. Astron. Astrophys.}}
\def\icarus{\ref@jnl{Icarus}}           
\def\jcap{\ref@jnl{J. Cosmology Astropart. Phys.}}
\def\jrasc{\ref@jnl{JRASC}}             
\def\memras{\ref@jnl{MmRAS}}            
\def\mnras{\ref@jnl{MNRAS}}             
\def\na{\ref@jnl{New A}}                
\def\nar{\ref@jnl{New A Rev.}}          
\def\pra{\ref@jnl{Phys.~Rev.~A}}        
\def\prb{\ref@jnl{Phys.~Rev.~B}}        
\def\prc{\ref@jnl{Phys.~Rev.~C}}        
\def\prd{\ref@jnl{Phys.~Rev.~D}}        
\def\pre{\ref@jnl{Phys.~Rev.~E}}        
\def\prl{\ref@jnl{Phys.~Rev.~Lett.}}    
\def\pasa{\ref@jnl{PASA}}               
\def\pasp{\ref@jnl{PASP}}               
\def\pasj{\ref@jnl{PASJ}}               
\def\rmxaa{\ref@jnl{Rev. Mexicana Astron. Astrofis.}}%
\def\qjras{\ref@jnl{QJRAS}}             
\def\skytel{\ref@jnl{S\&T}}             
\def\solphys{\ref@jnl{Sol.~Phys.}}      
\def\sovast{\ref@jnl{Soviet~Ast.}}      
\def\ssr{\ref@jnl{Space~Sci.~Rev.}}     
\def\zap{\ref@jnl{ZAp}}                 
\def\nat{\ref@jnl{Nature}}              
\def\iaucirc{\ref@jnl{IAU~Circ.}}       
\def\aplett{\ref@jnl{Astrophys.~Lett.}} 
\def\apspr{\ref@jnl{Astrophys.~Space~Phys.~Res.}}
\def\bain{\ref@jnl{Bull.~Astron.~Inst.~Netherlands}} 
\def\fcp{\ref@jnl{Fund.~Cosmic~Phys.}}  
\def\gca{\ref@jnl{Geochim.~Cosmochim.~Acta}}   
\def\grl{\ref@jnl{Geophys.~Res.~Lett.}} 
\def\jcp{\ref@jnl{J.~Chem.~Phys.}}      
\def\jgr{\ref@jnl{J.~Geophys.~Res.}}    
\def\jqsrt{\ref@jnl{J.~Quant.~Spec.~Radiat.~Transf.}}
\def\memsai{\ref@jnl{Mem.~Soc.~Astron.~Italiana}}
\def\nphysa{\ref@jnl{Nucl.~Phys.~A}}   
\def\physrep{\ref@jnl{Phys.~Rep.}}   
\def\physscr{\ref@jnl{Phys.~Scr}}   
\def\planss{\ref@jnl{Planet.~Space~Sci.}}   
\def\procspie{\ref@jnl{Proc.~SPIE}}   

\makeatother														

\usepackage{wrapfig}
\usepackage{rotating}
\newcommand{\figheight}{80mm}
\def\farcs{\hbox{$.\!\!^{\prime\prime}$}}

\begin{document}

\acrodef{sst}[SST]{Swedish 1-m Solar Telescope}
\acrodef{sdo}[SDO]{Solar Dynamics Observatory}
\acrodef{hmi}[HMI]{Helioseismic and Magnetic Imager}
\acrodef{fov}[FOV]{Field of View}
\acrodef{idl}[IDL]{Interactive Data Language}


\begin{titlepage}
\includegraphics[width=0.15\textwidth]{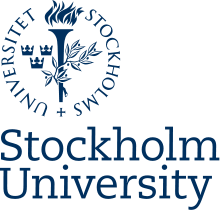}

\begin{center}
\textsc{\Large Bachelor's Thesis}

\vspace{1.5cm}
{\huge \bfseries Calibration of the SST Image Scale Through the Use of Imaging Techniques }

\vspace{1.5cm}
\begin{minipage}{0.4\textwidth}
	\begin{flushleft}
	\large \emph{Author:}\\
	Alexander Norén
	\end{flushleft}
\end{minipage}
\begin{minipage}{0.4\textwidth}
	\begin{flushright}
	\large \emph{Supervisor:} \\
	Mats Löfdahl
	\end{flushright}
\end{minipage}

\vfill
{\large August 22, 2013}

\vfill
{\large \emph{Submitted to:} \\ Peter Lundqvist  \\ Dan Kiselman}

\vspace{1.5cm}

Stockholm University \\ Department of Astronomy \\ AlbaNova University Center \\ Stockholm, Sweden

\end{center}
\end{titlepage}

\abstract{\noindent The \ac{sst} offers excellent imaging quality, but it has a comparatively small field of view. This means that while observing the solar photosphere, there has been no convenient way of calibrating the image scale of the telescope. Other telescopes, such as NASA's \ac{sdo} utilize their larger \ac{fov} to use the solar disk as a reference in order to measure the image scale.

In the past, the image scale of the \ac{sst} has been determined by measuring the distance between the moons of Jupiter in a captured \ac{sst} image and comparing it to reference values, as well as with the Venus transit of 2004. Both of these methods have their drawbacks, including needing to open the telescope at night or waiting for the very rare occurrence of a solar transit, which the telescope may not even be in a position to observe. Additionally, assessing the accuracy of these methods can be difficult. The purpose of this thesis is to examine the feasibility of an idea proposed by the faculty of the Institute for Solar Physics at Stockholm University, that would allow us to routinely calibrate the image scale of the \ac{sst} when desired and with known accuracy of the measurement, without the need to open the telescope at night.

\vspace{\baselineskip}\noindent
The measurements performed so far are consistent with the old value to about one third of a percent, with a total uncertainty of the SST/CRISP image scale of $\lesssim 0.1\%$. Resulting in a grid spacing of the pinhole array of $5\farcs15$, which can be used to determine the image scale of all the remaining science cameras of the \ac{sst}.

\newpage
\tableofcontents
\newpage
\listoffigures
\listoftables

\acresetall 
\newpage
\section{Introduction}

The Island of La Palma in the Canary Islands is home to the \ac{sst}. Which, with its unrivaled resolution for a ground-based solar telescope has become the worlds premiere source for high-resolution observations of the solar photosphere and chromosphere.

The goal of the \ac{sst} mission is to answer questions concerning such phenomena as solar magnetic fields, formation of stellar spectra and the dynamics of the upper solar atmosphere. However, despite the excellent resolution of the telescope and its ability to observe and photograph solar details with much greater fidelity than previously possible, there is a problem: the telescope has a very narrow \ac{fov} compared to the solar disc, which means we need some other reference when calibrating the image scale of the telescope. In the past, the image scale has been determined by observing the Venus disc during solar transits, as well as measuring the distance between the moons of Jupiter, whose orbits are known with high accuracy. Both methods do however have their drawbacks; Taking images of Jupiter's moons require night-time operations, which is undesirable and can also be dangerous to the personnel on site, while using the transit method require observations to be made during specific events and as such the interval between calibrations is limited to when these transits occur. Additionally, when observing the Venus transit, there is an added complication in form of Venus atmosphere; in that the atmosphere, when viewed from a distance, makes it difficult to determine the true diameter of the planet, which directly influences the accuracy of the discerned image scale. To this end, the goal for this thesis is to investigate the feasibility of an idea proposed the faculty of the Institute for Solar Physics, AlbaNova University Center, Stockholm, Sweden, which would allow the image scale of the \ac{sst} to be calibrated routinely, while still retaining, or perhaps even surpassing the accuracy of the transit method.

\vspace{\baselineskip}\noindent
The idea is to use continuum images of high-contrast structures on the photosphere of the Sun, taken by the \ac{sst} and then compare these with images taken by the \ac{hmi} instrument of the \ac{sdo} at approximately the same time. Since the \ac{hmi} has a well determined image scale, this would allow one to use imaging techniques in order to minimize the intensity difference between the two images over these structures, so that they might be fitted in a least-square sense. It should be noted that technically this does not provide a direct measure for the image scale of the \ac{sst} as a whole. What it does provide is the image scale in the specific camera used (of which there are several). During observations, images are taken with every science-camera of a so-called pinhole-array that is located at the focus present after the light leaves the telescope tube. These holes are distributed with regular distances from each other and serve as an unchanging reference. This is necessary as the image scale in a certain camera may change if the setup on the optical table is altered even slightly. And since the different cameras have varying image scales because of the different pixel-sizes and optics. By instead measuring the distance in pixels between the pinholes for the same camera and same occasion as the images used in determining the \ac{sst} image scale, it is possible to convert the distance between the holes to arc seconds and thereby determine the image scale for every camera simultaneously.

\subsection{The 1-m Swedish Solar Telescope}

\begin{figure}[h]
	\centering
	\includegraphics[width=0.45\textwidth, height=\figheight]{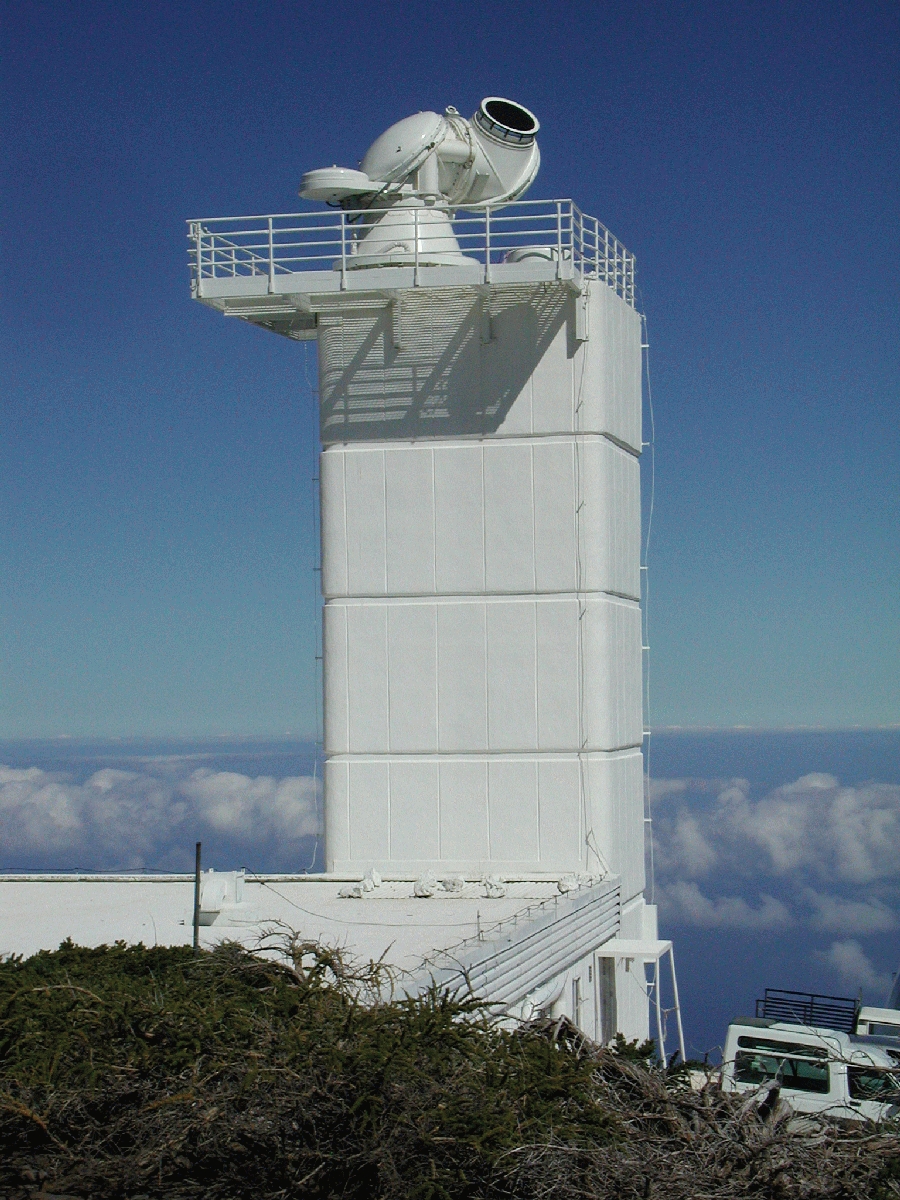}
	\includegraphics[width=0.45\textwidth, height=\figheight]{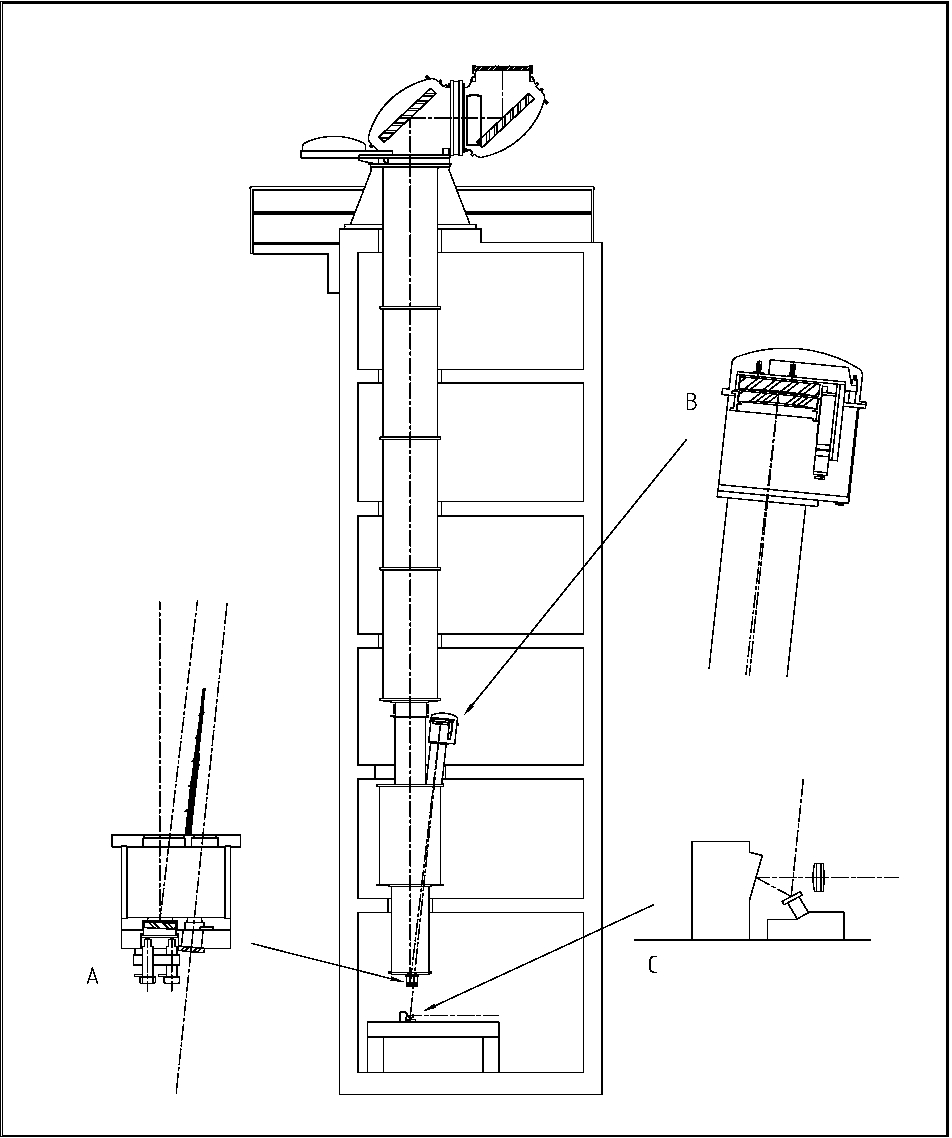}
	\caption[The Swedish 1-m Solar Telescope]{The tower supporting the Swedish 1-m Solar Telescope on LaPalma.\footnotemark \\ A.~Details of the
	box holding the field mirror and field lens. \\ B.~The Schupmann corrector with one lens and one mirror.  \\ C.~The re-imaging optics, located
	on the optical table and consisting of a tip-tilt mirror, adaptive optics and a re-imaging lens.}
	\label{image:sst_pictures}
\end{figure}
\footnotetext{Image credit:~\href{http://www.solarphysics.kva.se/}{The Institute for Solar Physics}}

\noindent
The Swedish 1-m Solar Telescope (figure~\ref{image:sst_pictures}) went live on March 2, 2002, and is the successor to the 47.5 cm Swedish Vacuum Solar Telescope, that was retired on the 28th of august, 2000. Located at the same position, at Roque de los Muchachos Observatory, La Palma, Spain. The \ac{sst} has a clear aperture of \SI{98}{\centi\meter}, more than double that of its predecessor, making it one of the largest optical solar telescope in the World. The \ac{sst} is run and operated by the Institute for Solar Physics, associated with the Department of Astronomy at Stockholm University.
\vspace{\baselineskip}

\noindent
All ground-based telescopes and observatories have one thing in common: they must all contend with imaging errors, in the form of \textit{seeing} (blurring caused by distortions and turbulence) that arises in the Earth's atmosphere. But with solar telescopes, there are added problems associated with the immense heat coming from the Sun. These problems manifest themselves, for example, through \textit{local seeing}. Local seeing is pretty much the same effect as regular seeing, only it happens because the Sun heats up the ground and surrounding area of the telescope. This causes 'heat waves' to rise and fluctuate in front of the lens, which can further compound the imaging errors due to deformation of the wavefront. Furthermore, the air inside of the telescope tube is also subject to heating, which causes more turbulence and degrades the resolution.

The \ac{sst} counteracts the latter effect by utilizing a vacuum-telescope design, thereby resolving many of the problems due to heating of the apparatus. The telescope is also unique in that it was the first solar telescope designed for use with \textit{adaptive optics} (upgraded in April, 2013). An adaptive optics system has a deformable mirror that can correct for optical aberrations and atmospheric wavefronts. In the case of the \ac{sst}, the mirror is able to adjust for the  blurring caused by the Earth's atmosphere up to $2000$ times per second. Such a system means you can get away with using fewer optical surfaces (mirrors, lenses), which always reduces the intensity a bit and may introduce refractive errors.

These two factors, together with the fact that the \ac{sst} uses a single 1-meter diameter glass lens to seal off the vacuum, instead of a flat vacuum window, as some others have done previously (e.g. Dunn Solar Telescope, Sacramento Peak, New Mexico), ensures excellent image quality. Furthermore, though the typical method of operation is to use filters to isolate narrow wavelength bands, whenever the need arises to observe in broader wavelengths, the image quality may begin to suffer because of chromatic aberrations (distortion caused because the lens have different refractive indexes for different wavelengths, hence different focal points). The \ac{sst} adjusts for this by employing a so-called Schupmann corrector (see schematic in figure~\ref{image:sst_pictures}), which redirects the light and effectively puts the different wavelengths into a single focus \citep{2003ASPC..307....3S}. This feature also comes in handy when utilizing the onboard spectograph and while changing wavelengths for narrow-band cameras, as it does away with the need to adjust the focus.

\subsection{The Transit Method}

\begin{figure}[h]
	\centering
	\includegraphics[width=0.41055\textwidth]{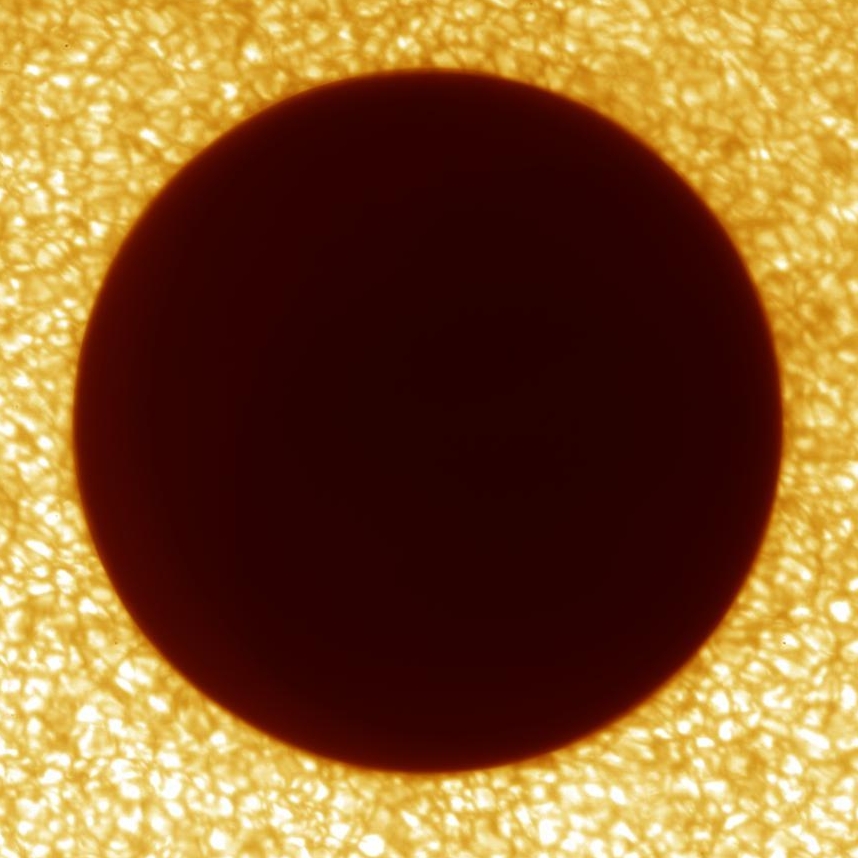}
	\includegraphics[width=0.4785\textwidth]{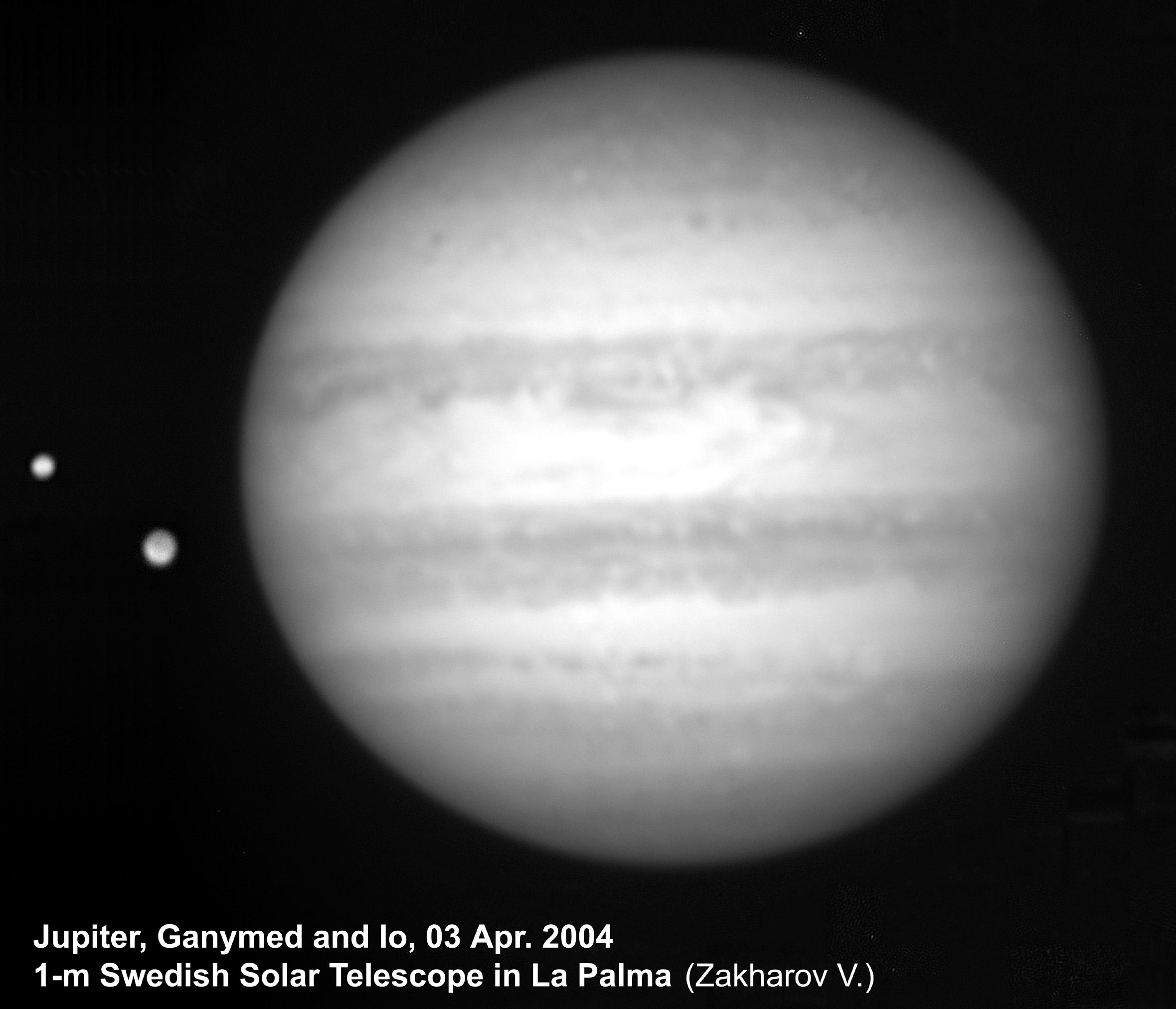}
	\caption[Venus transit \& Jupiter with Ganymede and Io]{Left: Venus transition across the Sun. Right: Jupiter with Ganymede and Io
	\footnotemark}
	\label{image:transits}
\end{figure}
\footnotetext{Image credit:~\href{http://www.solarphysics.kva.se/}{The Institute for Solar Physics}}

\noindent
As mentioned, the previous measurement of the \ac{sst} image scale was performed using the transit method. This entails measuring the relative size of an object transitioning in front of another at some zoom-factor of the telescope. The measurements are then traced back to a so-called pinhole array (figure~\ref{image:pinhole_array}) that is placed at the prime focus of the \ac{sst} at each day before the data collecting begins. Subsequently, the pinhole array may be used as an unchanging reference with which to calibrate the cameras, since the distance between the holes are static.

\vspace{\baselineskip}\noindent
The transit method was most recently employed during the Venus transit of 2004, and gave the currently used image scale of $0.0592$ arcsec/pixel in the CRISP camera.

The huge drawback of this method, other than the difficulties associated with the atmosphere of Venus and difficulties in assessing the accuracy in the measurements, are that Venus transits are very rare. In fact, other than the Venus transit that took place in June 2012 (which was not even visible from La Palma!), the next will not occur until the year 2117~\citep{2012A&A...547A..22G}.

\begin{figure}[h]
	\centering
	\includegraphics[height=\figheight]{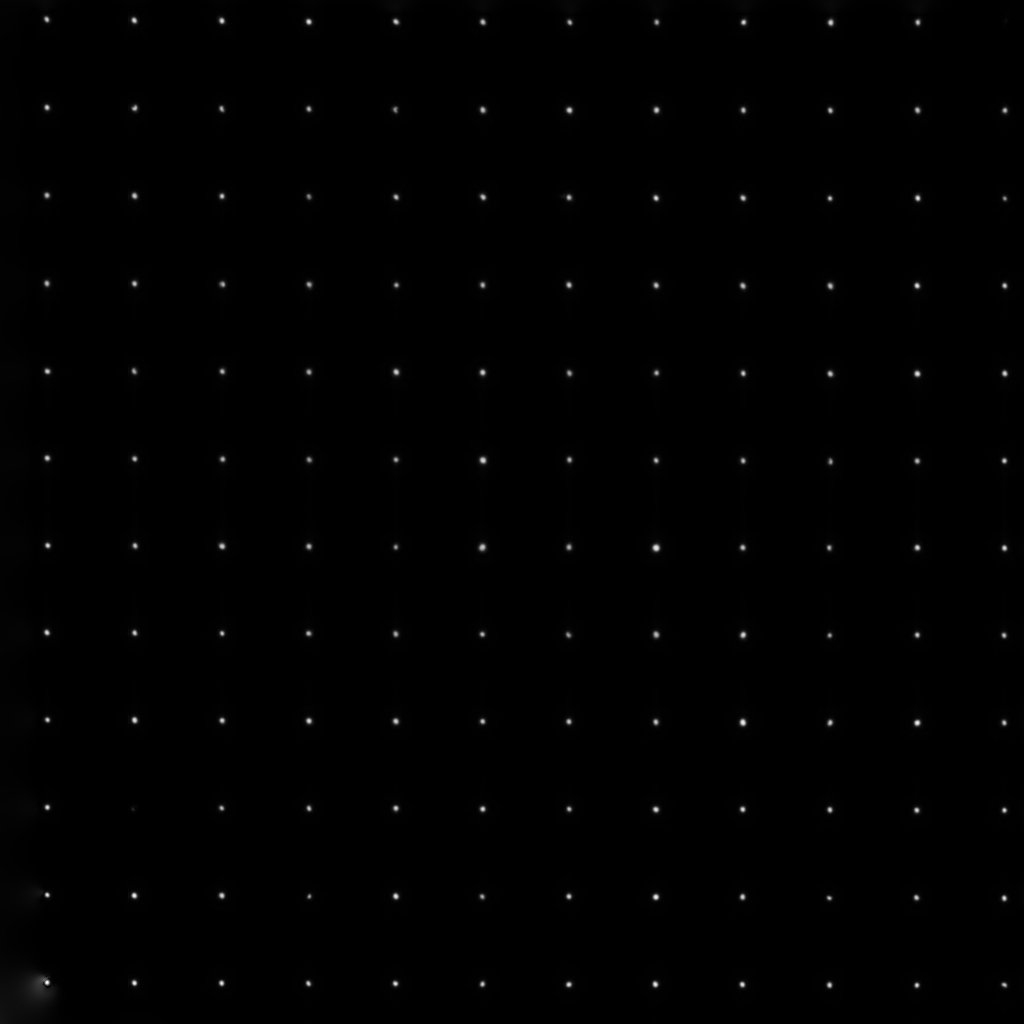}
	\caption[Pinhole array]{An image capture of the pinhole array at the primary focus of the \ac{sst} which is used as a fixed reference point
	to determine the image scale of the science cameras.
	\footnotemark}
	\label{image:pinhole_array}
\end{figure}
\footnotetext{Image credit:~\href{http://www.solarphysics.kva.se/}{The Institute for Solar Physics}}

\subsection{The Solar Dynamics Observatory}

The \ac{sdo} is a \textbf{NASA}/Goddard Space Flight Center mission launched on February 11, 2010, and is orbiting the Earth in a geosynchronous orbit. The system itself includes three instruments: the Helioseismic and Magnetic Imager (HMI), the Atmospheric Imaging Assembly (ATA) as well as the Extreme Ultraviolet Variablity Experiment (EVE). 
Unless otherwise cited, all information regarding the \ac{sdo} comes from~\citep{2012SoPh..275..229S}

\subsubsection{The Helioseismic and Magnetic Imager}

The instrument that will be of interest for this thesis, is the Helioseismic and Magnetic Imager (see figure~\ref{image:HMI_instrument}), which provides full-disk solar imaging at the Fe I \SI{6173}{\angstrom} line, using a \SI{14.0}{\centi\meter} aperture with an optical resolution that is better than $1$ arcsecond. It is designed to study oscillations and the magnetic field at the photosphere.

\begin{figure}[!h]
	\centering
	\includegraphics[height=\figheight]{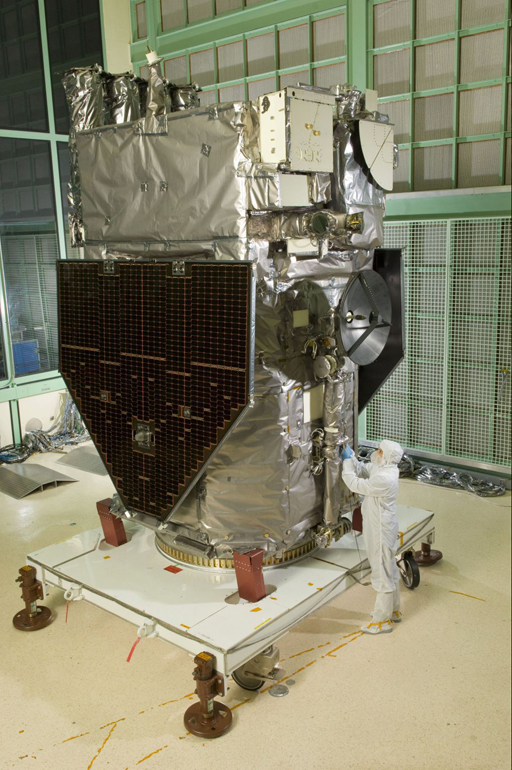}
	\includegraphics[height=\figheight]{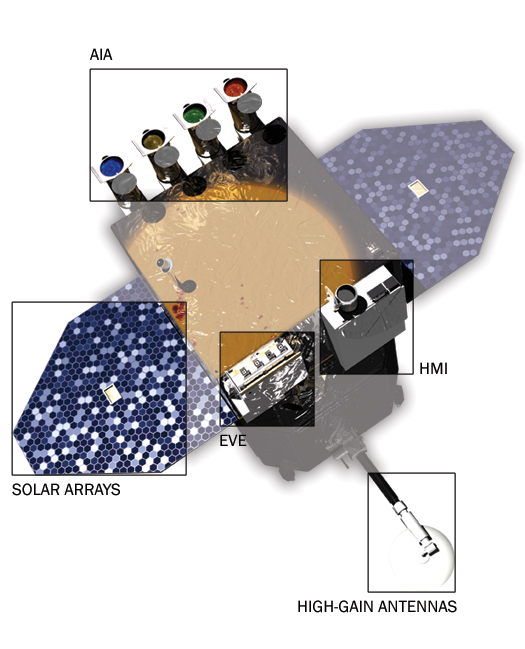}
	\caption[Photo of the \ac{sdo}]{The Solar Dynamics Observatory.
	\footnotemark}
	\label{image:SDO}
\end{figure}
\footnotetext{Image credit:~\href{http://sdo.gsfc.nasa.gov/mission/instruments.php}{NASA - Goddard Space Flight Center}.}

\begin{figure}[!h]
	\centering
	\includegraphics[height=\figheight]{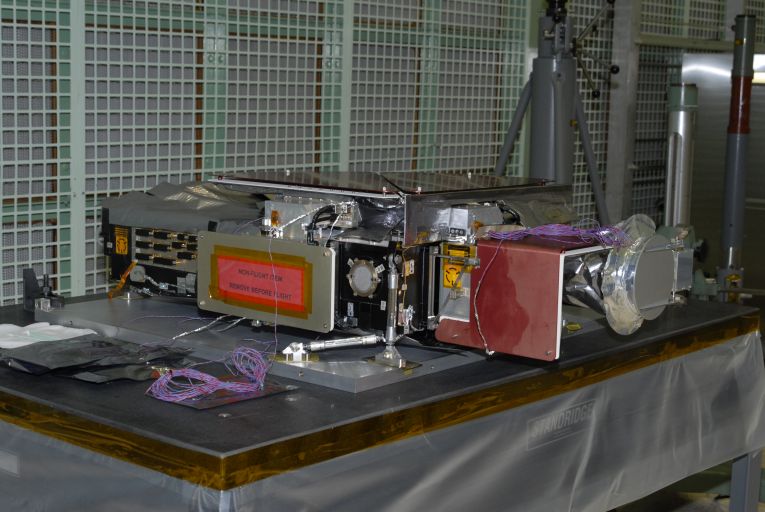}
	\caption[The Helioseismic and Magnetic Imager]{The HMI Instrument
	\footnotemark[\value{footnote}]}
	\label{image:HMI_instrument}
\end{figure}

\subsubsection{The HMI image scale}

Knowing the image scale of the \ac{hmi} and the limitations associated with its determination is important since this will be the basis for the resulting \ac{sst} image scale calculations. The way the image scale is calculated for the \ac{hmi} instrument is through the image geometry. More specifically, the limb of the Sun is measured as a function of wavelength to obtain the continuum value $R$. From this you get the image scale $\mathrm{CDELTX} = R_{\mathrm{obs}} / R$, where $R_{\mathrm{obs}} = \arcsin \left( R_{\mathrm{ref}} / D_{\mathrm{obs}} \right)$. In this context, $\mathrm{CDELTX}$ is the variable image scale of the \ac{hmi} present in the header file of each image, corrected for fluctuations in temperature (which influences the instrumentation) and variations in the distance to the Sun, caused by movement of both the Earth and the \ac{sdo} in their orbits.

\vspace{\baselineskip}\noindent
Explanation of notations:\\
$R$ (solar radius in pixels)\\
$R_{\mathrm{obs}}$ (solar radius in arcsec)\\
$R_{\mathrm{ref}}$ (assumed solar radius in m)\\
$D_{\mathrm{obs}}$ (distance to the sun in m)

\vspace{\baselineskip}\noindent
Technically, this does not give the image scale at the disk center, but rather the mean image scale from center to limb. Though the error is negligible compared to the uncertainty of $R$ and $R_{\mathrm{ref}}$\citep{schou13}.

A problem is what value to use for $R_{\mathrm{ref}}$. As of February, 2013 the value used by the \ac{hmi}-team was \SI{696}{\mega\meter}, with a resulting (slightly varying) image scale of $\sim 0.505$ arcsec/pixel. But this value for $R_{\mathrm{ref}}$ is still a matter of some controversy, and others have gotten different values, ex. \SI{695.508 \pm 0.0256}{\mega\meter} \citep{1998ApJ...500L.195B}. Comparing the numbers cited we see that the difference is roughly $0.07\%$. Implying, if we ignore the arcsine because of the small angle, that $\mathrm{CDELTX} = R_{\mathrm{ref}} / \left( R D_{\mathrm{obs}} \right)$, which should then have about the same relative errors as $R_{\mathrm{ref}}$. Suffice to say, it is the value that has the inherently largest uncertainty associated with it, and these errors are carried over to the image scale determination of the \ac{hmi}, when using the method suggested in this project. Which is something to keep in mind.

The good news is that as this value is further refined by the \ac{hmi} team, so too will the value of subsequent fittings be improved, since the current \ac{hmi} image scale is present in the headers of the image files provided.

\subsection{Programming}

The bulk of the work done during this project has been in the form of writing code in the \ac{idl}, for the purpose of automating the image scale determination process, so that future calibrations will be as effortless as possible. This includes, but is not restricted, to writing the actual code needed and adapting functions written in other languages (such as ANA) into \ac{idl}. Several library routines were also used, including some developed at the Institute for Solar Physics.

\vspace{\baselineskip}\noindent
During the course of the project, there were some changes to the initial plan as far as writing the actual code was concerned. It quickly became apparent that the least-squares method MPFIT was prone to getting stuck at local minima during the optimization process, when provided with initial guesses for an optimal fit that were too (but in the authors opinion, not unreasonably) far off. Which led to the examination of other possibilities. The solution that was settled upon was incorporating a grid-matching method as an initial fit to improve upon the guess before letting MPFIT do the final adjustments, thereby avoiding that pitfall and as a result ensuring more leeway in the initial parameter guess provided by the user before initializing the program.

\newpage
\section{Methods}

In order to determine an image scale for the \ac{sst}, a series of photographs depicting active regions (high-contrast structures) of the photosphere, spanning some time-interval (about 20 minutes in this case, divided over 40 image pairs) are compared between \ac{sst} and \ac{hmi} images of the same \ac{fov}. These images are then fitted as closely as possible using least-square methods by optimizing the four parameters: position (x,y), rotation and the zoom factor needed.

To achieve this, an \ac{idl} program was written, with the purpose of loading relevant \ac{sst} and \ac{hmi} images, recording the time and date that the images were taken and then minimize the time differential between each image. With the 30-second cadence of the \ac{sst} and 45-second cadence of the \ac{hmi}, this gives time matches better than $22.5$ seconds (average of $12.3$ s for the current run) in time. The time-matched images are then loaded as square matrices, where each element in the matrix represents the intensity of a pixel in the \ac{fov}. By minimizing the difference in intensity between each image pair a resulting zoom-factor is obtained. Thus, by dividing the \ac{hmi} image scale by this zoom-factor, you arrive at an calculated image scale of the \ac{sst}. Following this procedure for all the matched image pairs and taking the mean gives the resulting value of the \ac{sst} image scale, while the standard deviation of all pairs provides an uncertainty (not including any uncertainty added by the determination of the \ac{hmi} image scale).

\vspace{\baselineskip}\noindent
There are of course some difficulties with matching an \ac{hmi} image to an \ac{sst} image. The first and most obvious being that the image scale, \ac{fov} and resolution of the two instruments are very different (see figure~\ref{image:sst+hmi_original}).

\begin{figure}[!h]
	\centering
	\includegraphics[width=.45\textwidth]{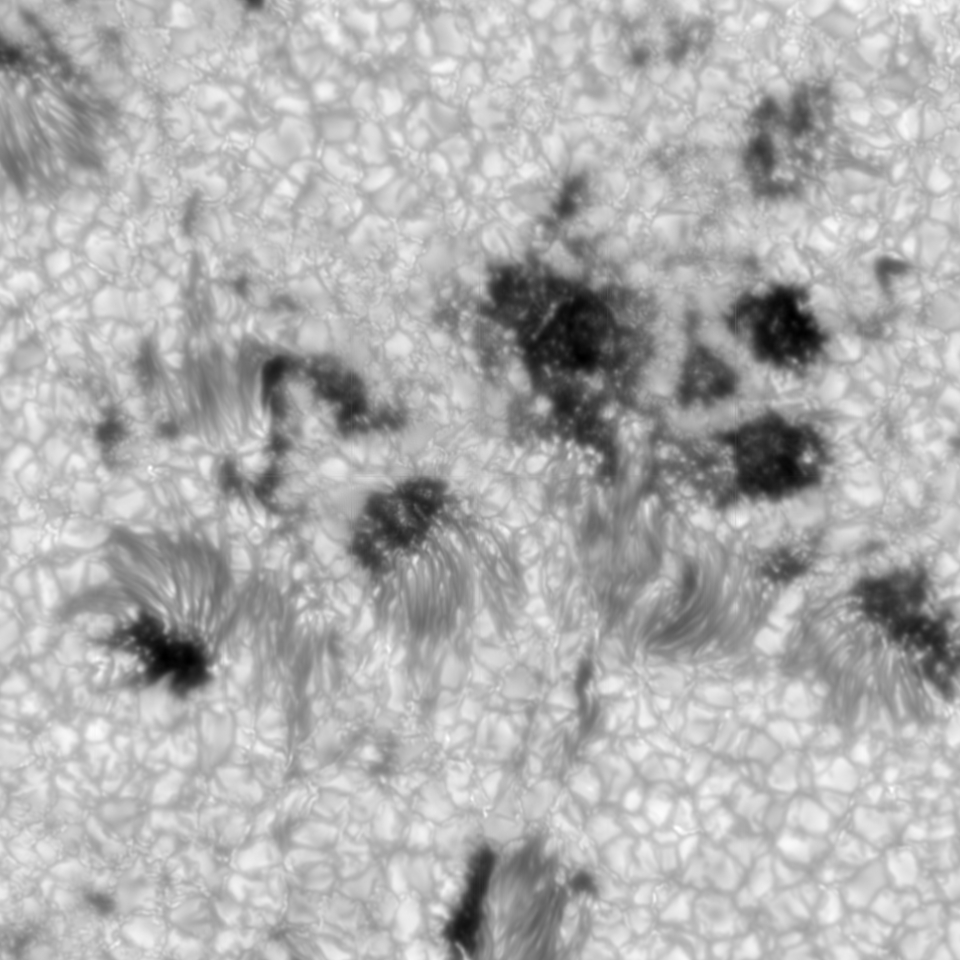}
	\includegraphics[width=.45\textwidth]{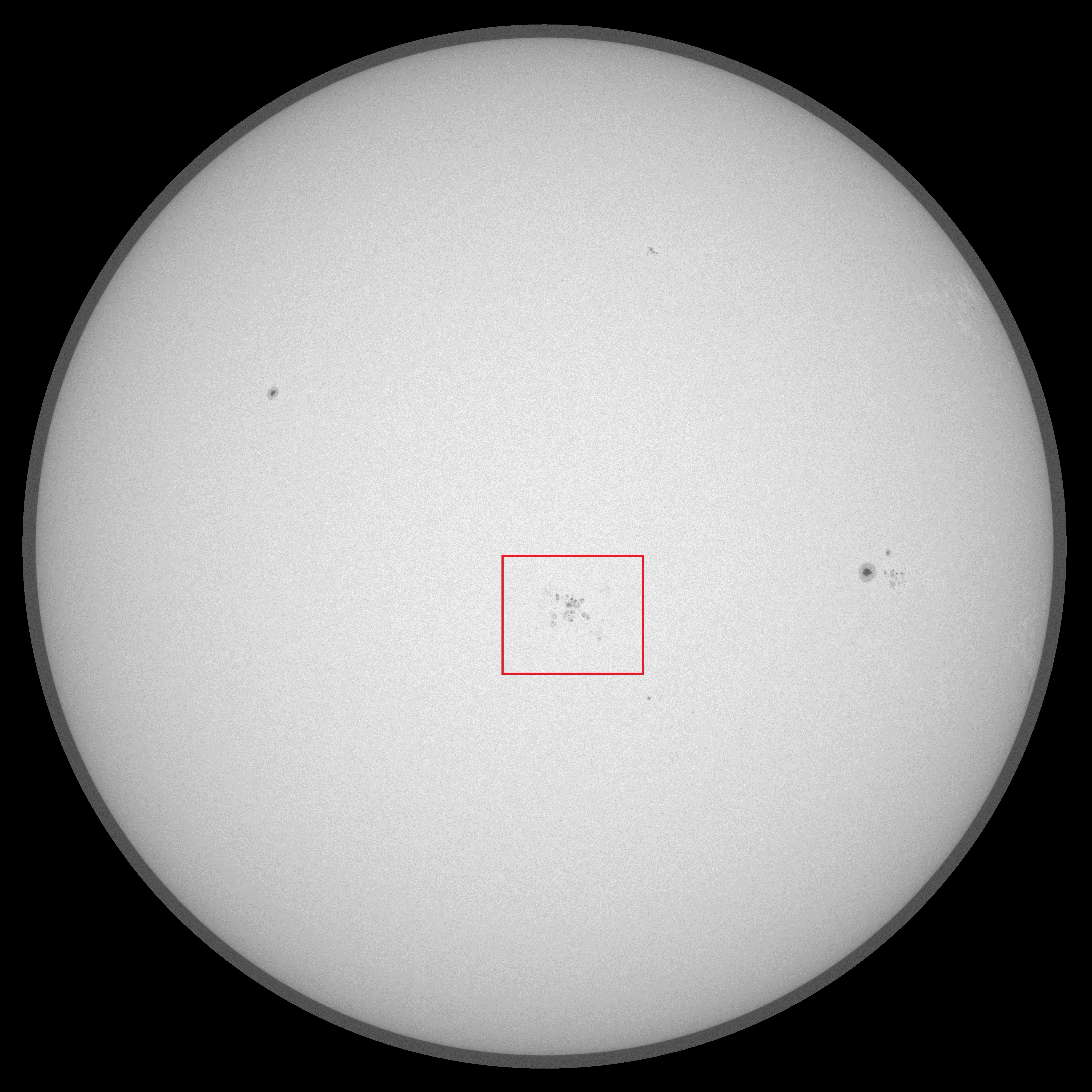}
	\caption[Original images taken by the SST \& HMI]{Images taken by the \ac{sst} and \ac{hmi} before any image processing. The red square in the
	\ac{hmi} image roughly corresponds to the area visible in the \ac{fov} of the \ac{sst}. Not including any any rotations that need to be made
	for the observed features to line up.}
	\label{image:sst+hmi_original}
\end{figure}

\subsection{Preparations}

To resolve these problems, the \ac{hmi} image is centered on the same area visible in the \ac{sst} image, but using a wider \ac{fov} to allow for corrections to subsequent parameter adjustments while still maintaining an image with a bare minimum of distortions caused by interpolation methods. This image is then used as a base for all succeeding image-processing. 

At this point, the \ac{hmi} base and \ac{sst} image shares roughly the same positioning and rotation. But there is still the problem of the significant difference in resolution between the two instruments, which causes the zoomed-in \ac{hmi} image to be very blurry in comparison to the higher resolution of the \ac{sst}. To counteract this, the \ac{sst} image is run through an IDL routine that degrades an image taken with a larger aperture telescope into the resolution of a smaller aperture telescope, taking into consideration any existing difference in observed wavelength.

With the \ac{sst} image degraded, the image pair should now be quite similar. Although, there is still one more thing that can be done before a fair comparison between them can be performed. This final step is to normalize the images, so that areas of similar contrast actually correspond in intensity over the images. This is accomplished by subtracting the mean value of the image and dividing it by the standard deviation. i.e.

\begin{verbatim}
IDL> sst_im_norm = (sst_im - mean(sst_im))/stdev(sst_im)
IDL> hmi_im_norm = (hmi_im - mean(hmi_im))/stdev(hmi_im)
\end{verbatim}

A comparison between the resulting images can be seen in figure~\ref{image:HMI_cropped}.

\begin{figure}[!h]
	\centering
	\includegraphics[width=.45\textwidth]{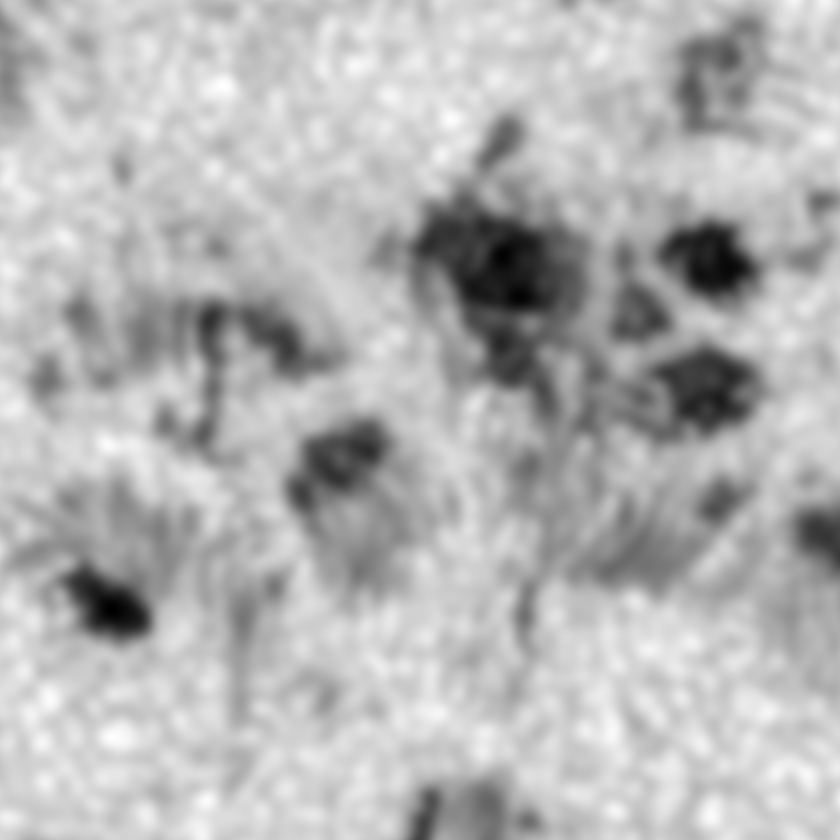}
	\includegraphics[width=.45\textwidth]{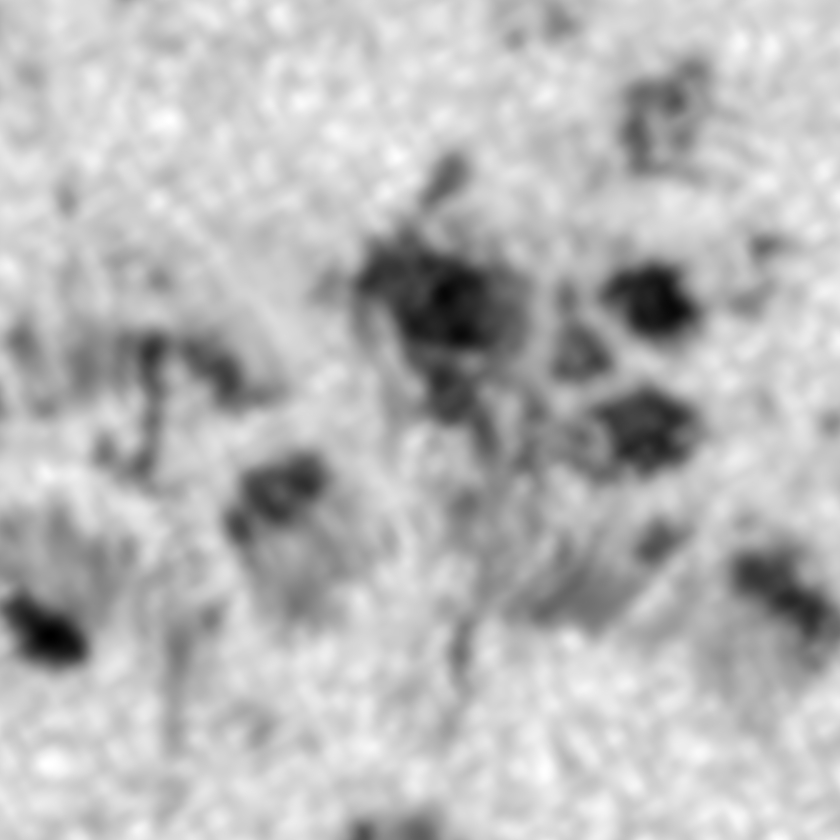}
	\caption[SST \& HMI comparison before fitting process]{On the left: the SST image, after being degraded \& normalized to roughly the same
	resolution as the \ac{hmi} image. On the right: the \ac{hmi} image of the cropped area contained within the red square in
	figure~\ref{image:sst+hmi_original}, before the fitting process has begun.}
	\label{image:HMI_cropped}
\end{figure}
\vspace{\baselineskip}

\subsection{The Fitting Process}

With all the preparations out of the way, the fitting process may finally commence. This is accomplished in two stages. Firstly, using a grid method called gridmatch\footnote{\url{http://ana.lmsal.com}} (originally implemented in the ANA programming language by Dr Richard Shine at the Lockheed Martin Solar and Astrophysics Laboratory~\citep{1994ApJ...430..413S}) as a rough preliminary fit and secondly through use of the non-linear least-squares fitting function MPFIT~\citep{2009ASPC..411..251M} for the final fine-tuning of the fitting parameters.

The grid method is a more coarse method than MPFIT for finding a fit between two images but it has the benefit of not being as prone to optimize the fit towards a local minimum rather than the global. This is beneficial as it provides more leeway in terms of the restriction on how closely the initial guess parameters must correspond to the actual values. When gridmatch has returned the improved parameters, the two resulting images should be very similar and MPFIT uses these parameters as an initial guess for its least-square fitting process, in order to refine the parameters at sub-pixel level accuracy.

\subsubsection{Grid Matching Method}

What this method does, is in principle to split the image that you are performing operations on (\ac{hmi}) into a grid, containing several sub-images and then try to fit these sub-images onto corresponding areas on the reference (\ac{sst}) image, by aligning the two images in terms of translations for every sub-image, while still allowing for some overlap for optimization. The routine then accumulates all these results from the various sub-images and returns the parameters (translation, rotation and zoom) that best correspond.

For the purposes of this project, gridmatch was set up so that the \ac{hmi} image was split into 2x2 sub-images, which were matched for a best-fit, allowing for a large overlap. The process is then iterated upon, but now the image is instead divided up into 4x4 sub-images. However this time with tighter restriction on the allowed overlap, as the fit should already be improved. This goes on progressively to a maximum of 16x16, or 256 sub-images (arbitrary limit set after experimentation with various image-pairs).

\subsubsection{MPFIT}

At this stage the parameters determined by the grid matching method, which should be very good, are passed on as an initial guess to MPFIT. And the function may proceed to optimize the fit even further by performing the \textit{Levenberg-Marquardt algorithm}, also known as the \textit{damped least-squares} method, in order to minimize the difference in intensity between the two images. Until a final, best-fit value for all parameters has been determined. Though, at this point the only parameter of any real interest is, of course, the zoom-factor. As this allows one to determine the image scale simply by dividing the \ac{hmi} image scale by the determined zoom-factor. The resulting best-fit comparison can be seen in figure~\ref{image:final_fit}.

\begin{figure}[!h]
	\centering
	\includegraphics[width=.45\textwidth]{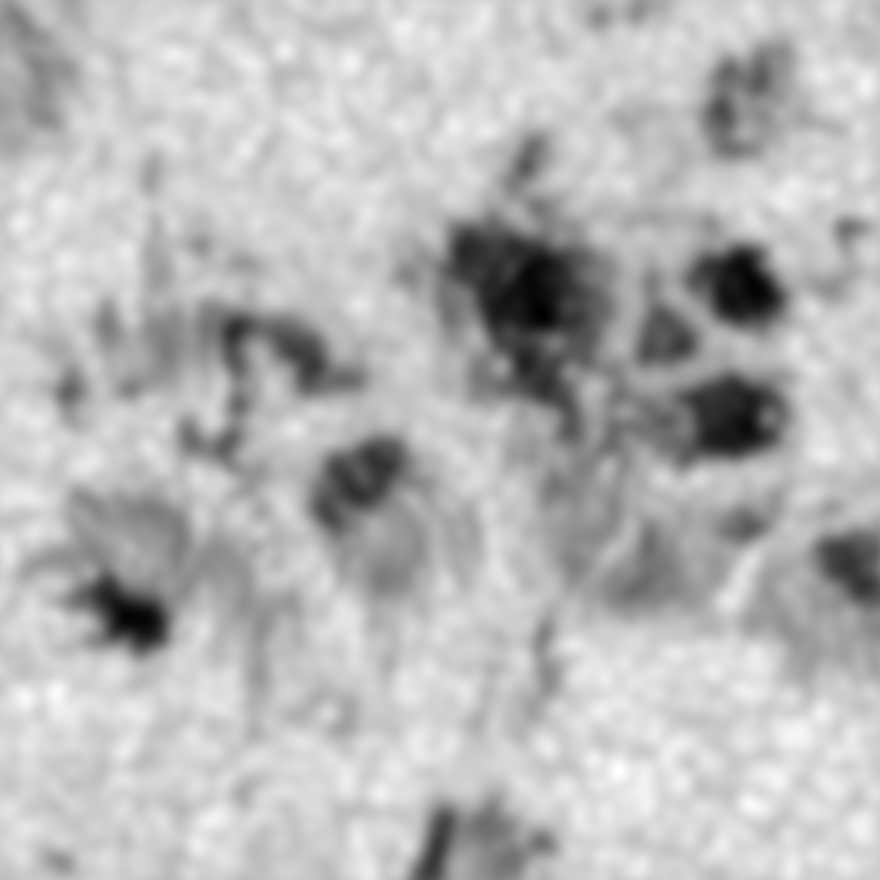}
	\includegraphics[width=.45\textwidth]{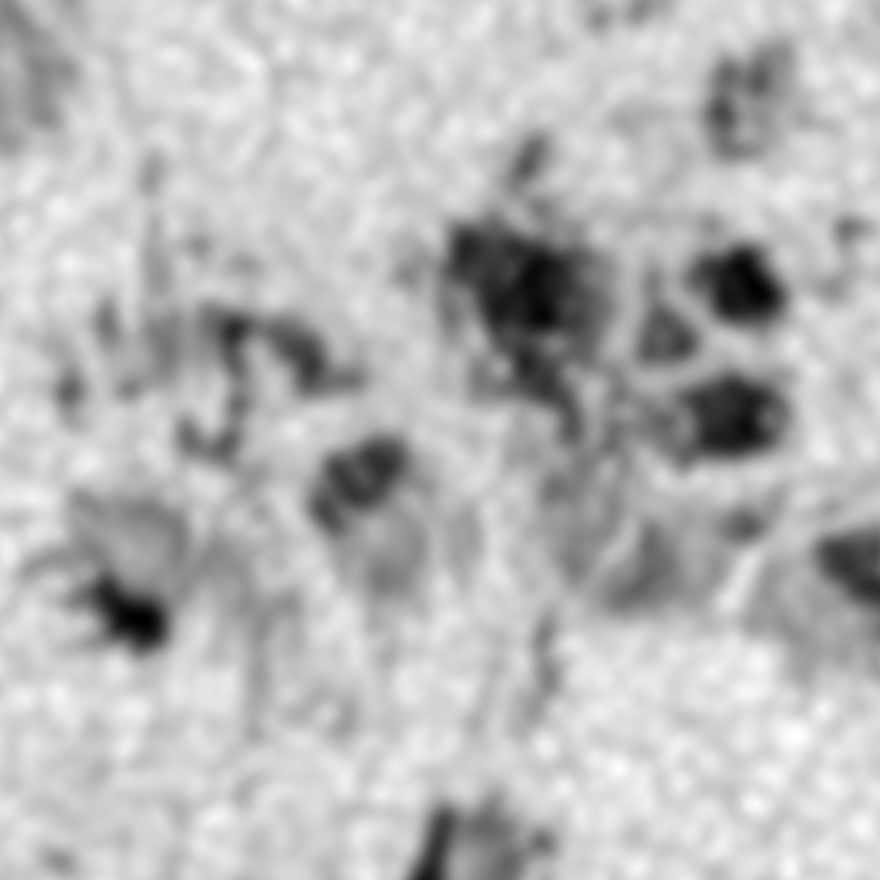}
	\caption[SST \& HMI comparison after fitting process]{On the left: the SST image, degraded and normalized. On the right: the HMI image
	after the fitting process has been completed.}
	\label{image:final_fit}
\end{figure}

\subsection{Accuracy of Fitting Processes}

As previously mentioned, the uncertainty of the fit is simply determined by means of the standard deviation across all used image-pairs (i.e. 40 for now).

\vspace{\baselineskip}\noindent
How accurately MPFIT is able to optimize the zoom-factor is of course of great concern when considering that a fit between \ac{hmi} and \ac{sst} will never be perfect. The telescopes are after all quite different and degrading the \ac{sst} image to a lower resolution and using interpolation methods on the \ac{hmi} image will always serve to impede how well a fitting can be performed. These factors directly influence just how carefully the zoom-factor can be optimized during the MPFIT process. Therefore, it is important to convince oneself that MPFIT actually is capable of optimizing the chi-square minimum to within the standard deviation if this uncertainty is to be taken seriously and the method to be considered viable.

\vspace{\baselineskip}\noindent
For this purpose, a routine was written which recreates image pairs from the fitting process and slightly varies only the zoom-factor as to provide data-points which may be used to examine whether a well behaved minimum is achievable within the uncertainty specified. Additionally, when viewing images illustrating the difference in intensity between the \ac{hmi} and \ac{sst} image fittings, there should be a clear tendency towards a more homogeneous surface after the gridmatch and MPFIT processes, as compared to before the optimization functions began.

\begin{figure}[!h]
	\includegraphics[width=0.32\textwidth]{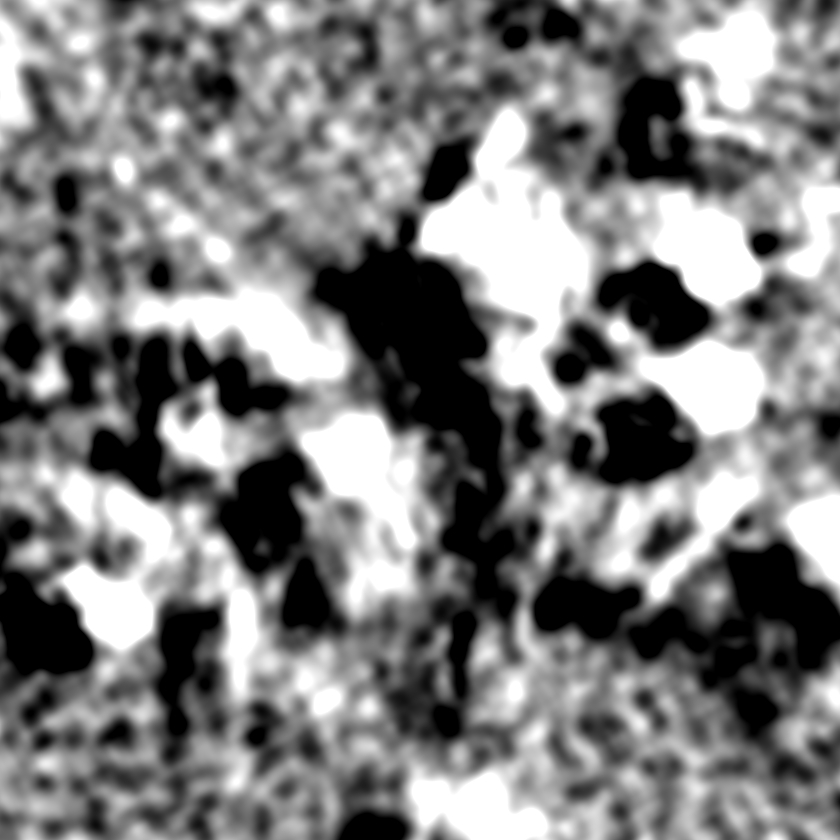}
	\includegraphics[width=0.32\textwidth]{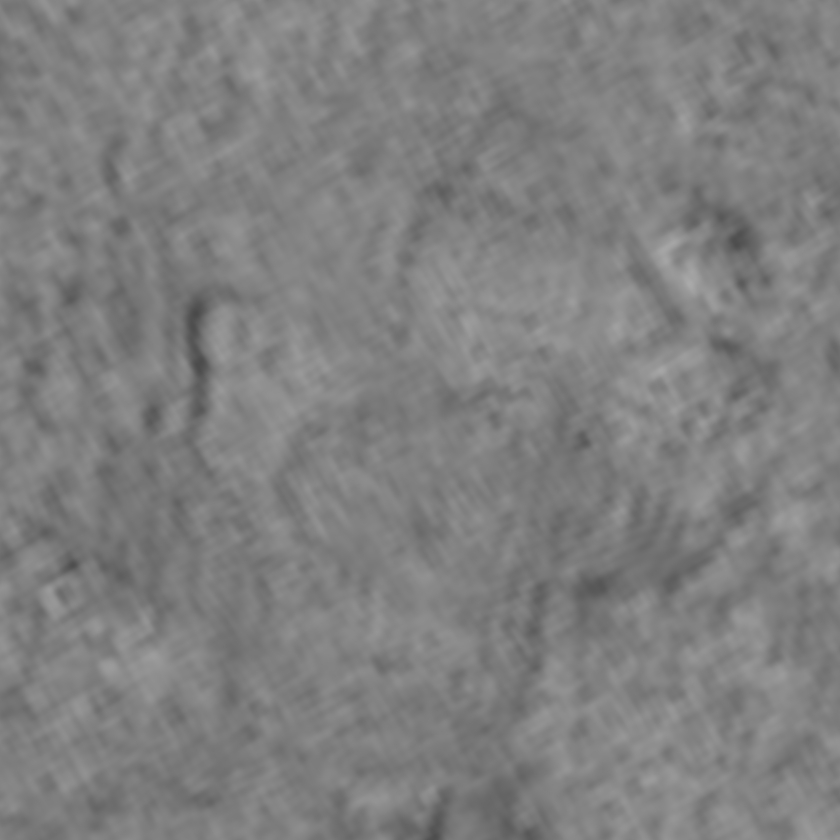}
	\includegraphics[width=0.32\textwidth]{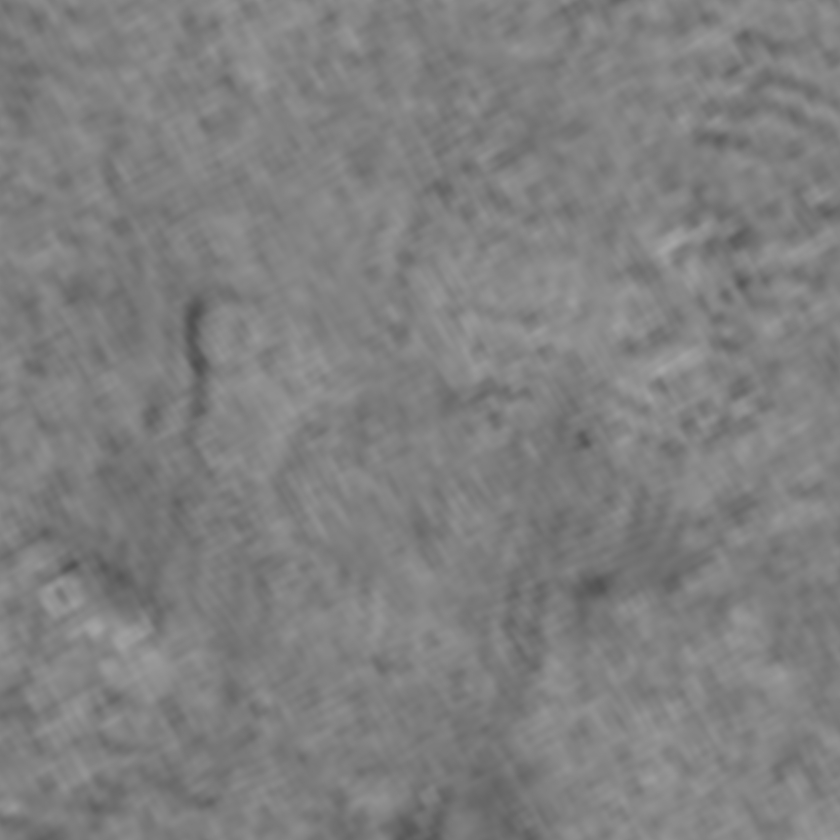}
	\caption[Difference images: initial, gridmatch \& MPFIT]{Difference between an \ac{hmi} and \ac{sst} images during the fitting
	process. Leftmost image: before the fitting process	has begun. Middle image: after the gridmatch process. Rightmost image: after the MPFIT
	process has completed.}
	\label{image:3waycomparison}
\end{figure}

\vspace{\baselineskip}\noindent
Figure~\ref{image:3waycomparison} illustrates a three-way comparison of the difference in intensity between the  the \ac{hmi} and \ac{sst} images before the fitting process has begun (but using the initial guess provided by the user), the difference after the gridmatch process has been completed and finally after the MPFIT adjustments.

The black and white features visible in the initial comparison highlights high-contrast areas of the two separate images that are not aligned correctly; hence causing deformations of areas with higher and lower intensity. The white spots are areas of high intensity and the black of lower. The fitting process adjusts the relative position between the images in order to merges these areas together, resulting in a grey-scaled image, that ideally would be completely homogeneous. In reality, some variations will always be present to some extent as the images are not taken at exactly the same instant and because of the great difference in resolution and relative positions of the telescopes. Even though the resolution difference is minimized through the degradation process of the \ac{sst} image, this process is not perfect and as such some variations are to be expected. All in all, the final fits are definitely adequate for our needs.

\newpage
\section{A More Detailed Description of the Program}

A rough breakdown of how the program works:

\vspace{\baselineskip}\noindent
First, the user provides the desired \ac{sst} and \ac{hmi} images taken during the same time interval in sub-folders within the program catalog. Secondly, because the \ac{fov} of the \ac{sst} and \ac{hmi} are very different, the program needs an initial parameter guess as to how the \ac{hmi} image should be manipulated in order to roughly correspond to the \ac{fov} of the \ac{sst} image (there is some room for error here, although the closer the initial parameters are, the better. Such an parameter guess is only needed for the very first, i.e. the earliest, image-pair of the current set). The fitting process may commence after modifying the user-input file and providing:

\begin{itemize}
	\item Initial coordinates (x,y), zoom-factor and rotational angle.
	\item the currently best known value of the SST image scale.
	\item the observed wavelength of the SST camera (the HMI wavelength is read from the header file).
\end{itemize}

\noindent
The program then: 

\begin{enumerate}
	\item Loads the images into arrays, where the value of each array cell corresponds to the intensity in a single pixel. The time-
	stamps of the images are also read from the file-names and image headers, whereby the program matches the \ac{sst} and \ac{hmi} images in pairs
	which have the shortest time differential between them. Unless there are gaps in the \ac{hmi} images provided, the cadence of the images should
	ensure that the time differential between them never exceeds 22.5 seconds. Should this limit be exceeded, a warning is displayed.

	\item Begins an iterative process over the image series provided, where a cutout (hmi\_base) is made from the \ac{hmi} image. Using the
	provided initial guess parameters the \ac{hmi} image is zoomed to approximately \ac{sst} image scale, but with a generous overhead in the
	\ac{fov}. This cutout serves as a reference image towards which parameter changes may be applied, including rotational changes. When these
	changes are done, the image can be cut to the same \ac{fov} as that of the \ac{sst} image without introducing imaging errors at the borders.
	
	\item Degrades the \ac{sst} image to the resolution of the hmi\_base image, as described in chapter 2.
	
	\item Runs the gridmatch subroutine, which divides the \ac{hmi} image into a grid pattern containing sub-images that are individually fitted
	against corresponding areas on the \ac{sst} image and determines the mean change needed in translation, rotation and zoom-factor in order to
	achieve a better fit than the initial guess parameters that were provided by the user.
		
	\item Passes the improved parameters on to the MPFIT subroutine, which then fine-tunes the parameters at a sub-pixel level to the best-fit
	values.	These are considered as good as it gets, and the final zoom-factor derived is used to calculate the image scale of the \ac{sst} image in
	this particular pair. This value is saved in a list that progressively fills up as the image pairs are processed.
\end{enumerate}

When the program has finished with one image pair the resulting best-fit parameters are sent on to the next image-pairs as a new initial guess (corrected for the trend of changes between previous image pairs when enough of them have been processed), which eliminates the need for user input on subsequent image pairs and speeds up the process since this new guess already is very close to optimal. When all of the provided image pairs have been processed, a final image scale is calculated from the mean. Which is displayed along with the standard deviation of the set and saved to a file. Giving the resulting image scale of the \ac{sst} camera used.

\newpage
\section{Results}

\subsection{Calculated Image Scale}

Processing 40 pairs of SST/CRISP and SDO/HMI images of AR1589, collected near the disk center on 2012-10-15 between 09:30 and 09:50 resulted in a calculated image scale of $0.058994 \pm 4.5 \times 10^{-5}$ arcsec/pixel. Below is an excerpt from the program output.

\begin{verbatim}
Finished:       40 image pairs
---------------------------------------------
FINAL RESULTS FROM ALL IMAGES:
resulting SST arcsec per pix:	 0.058993879 
factor (new/old):       		      0.99651822
change from previous value:		  -0.00020612141

Standard deviation from mean:  4.5488712e-05
Maximum deviation from mean:   0.00013081031
Minimum deviation from mean:   3.7977090e-07
----------------------------------------------
\end{verbatim}

\noindent
In comparison to the value used since 2004 of $0.0592$ arcsec/pixel, this constitutes a revision of the image scale of the \ac{sst}/CRISP camera by $\sim 0.35\%$. More specific data, including the calculated image scale for each separate pair, is available in table~ \ref{table:images_and_results}, which shows the time difference between the \ac{hmi} and \ac{sst} image as well as the calculated image scale and change from the previous value in milliarcsecond (mas)/pixel for each of the forty image pairs.

\vspace{\baselineskip}\noindent
Since the fitting process employed to achieve the final image scale was done using two separate methods; the gridmatching method as a coarse means of improving the initial guess and MPFIT to fine-tune the parameters obtained through the gridmatching, it may be instructive to examine just how much these methods actually differ in accuracy. Calculating the image scale and uncertainty by the mean and standard deviation of the distribution, while using the respective zoom-parameters of gridmatch and MPFIT gave the results shown in table~\ref{table:imscale_methods}.

\begin{table}[h!]
\centering
\begin{tabular}{lcc}
\toprule\toprule
Method & image scale (mas/pixel) & uncertainty (\%) \\
\toprule
V. Transit 		& $59.2 \pm $ n/a		& n/a		\\
Gridmatch 		& $59.00 \pm 0.61$		& 0.01034	\\
MPFIT			& $58.994 \pm 0.045$	& 0.00077
\end{tabular}
\caption[\ac{sst}/CRISP image scale using different methods]{\ac{sst}/CRISP image scales (in milliarcsec (mas)/pixel) calculated using the Venus transit as well as only through use of the gridmatch method and with the improved image scale determined through MPFIT including uncertainties.}
\label{table:imscale_methods}
\end{table}

\noindent
Table~\ref{table:imscale_methods} shows that while both methods are functional and do arrive at a similar image scale; the uncertainty achieved through gridmatch is larger than the uncertainty of MPFIT by more than a factor of $13$. Giving credence to using MPFIT to fine-tune the parameters. \textit{note: all \ac{sst} measurements were performed through the \SI{5576}{\angstrom} CRISP prefilter.}

\begin{table}[p]
\centering
\resizebox{\textwidth}{!}{\begin{minipage}{\textwidth}
\caption[Timestamps \& resulting image scales]{Timestamps with resulting values and changes for the calculations of each image pair. \textit{Note: all photographs were taken on the 15th of October, 2012}.}
\label{table:images_and_results}
\begin{tabular}{lcccccc}
\toprule\toprule
index & SST image & HMI image & $\Delta t$ & image scale & new/old & change\\
& timestamp & timestamp & (sec) & (mas/pixel) & & (mas/pixel)\\
\toprule
0   & 09:30:24  &  09:30:45  &  21  & $59.03163$ &  $0.99716$  &  $-0.00017$\\
1   & 09:30:53  &  09:30:45  &   8  & $59.01287$ &  $0.99684$  &  $-0.00019$\\
2   & 09:31:23  &  09:31:30  &   7  & $59.02214$ &  $0.99700$  &  $-0.00018$\\
3   & 09:31:53  &  09:32:15  &  22  & $58.99497$ &  $0.99654$  &  $-0.00021$\\
4   & 09:32:23  &  09:32:15  &   8  & $59.00866$ &  $0.99677$  &  $-0.00019$\\
5   & 09:32:53  &  09:33:00  &   7  & $59.02839$ &  $0.99710$  &  $-0.00017$\\
6   & 09:33:23  &  09:33:45  &  22  & $58.96074$ &  $0.99596$  &  $-0.00024$\\
7   & 09:33:52  &  09:33:45  &   7  & $59.01839$ &  $0.99693$  &  $-0.00018$\\
8   & 09:34:22  &  09:34:30  &   8  & $58.97428$ &  $0.99619$  &  $-0.00023$\\
9   & 09:34:52  &  09:34:30  &  22  & $58.95655$ &  $0.99589$  &  $-0.00024$\\
10  & 09:35:22  &  09:35:15  &   7  & $59.00175$ &  $0.99665$  &  $-0.00020$\\
11  & 09:35:52  &  09:36:00  &   8  & $58.87797$ &  $0.99456$  &  $-0.00032$\\
12  & 09:36:22  &  09:36:00  &  22  & $58.89781$ &  $0.99490$  &  $-0.00030$\\
13  & 09:36:51  &  09:36:45  &   6  & $59.10221$ &  $0.99835$  &  $-0.00010$\\
14  & 09:37:21  &  09:37:30  &   9  & $58.97470$ &  $0.99619$  &  $-0.00023$\\
15  & 09:37:51  &  09:37:30  &  21  & $58.99599$ &  $0.99655$  &  $-0.00020$\\
16  & 09:38:21  &  09:38:15  &   6  & $58.94181$ &  $0.99564$  &  $-0.00026$\\
17  & 09:38:51  &  09:39:00  &   9  & $58.99271$ &  $0.99650$  &  $-0.00021$\\
18  & 09:39:21  &  09:39:00  &  21  & $58.95898$ &  $0.99593$  &  $-0.00024$\\
19  & 09:39:51  &  09:39:45  &   6  & $58.99600$ &  $0.99655$  &  $-0.00020$\\
20  & 09:40:21  &  09:40:30  &   9  & $59.02209$ &  $0.99699$  &  $-0.00018$\\
21  & 09:40:51  &  09:40:30  &  21  & $58.98419$ &  $0.99635$  &  $-0.00022$\\
22  & 09:41:21  &  09:41:15  &   6  & $58.97353$ &  $0.99617$  &  $-0.00023$\\
23  & 09:41:51  &  09:42:00  &   9  & $59.02195$ &  $0.99699$  &  $-0.00018$\\
24  & 09:42:21  &  09:42:00  &  21  & $58.92789$ &  $0.99540$  &  $-0.00027$\\
25  & 09:42:51  &  09:42:45  &   6  & $58.98710$ &  $0.99640$  &  $-0.00021$\\
26  & 09:43:21  &  09:43:30  &   9  & $59.00748$ &  $0.99675$  &  $-0.00019$\\
27  & 09:43:51  &  09:43:30  &  21  & $58.99350$ &  $0.99651$  &  $-0.00021$\\
28  & 09:44:21  &  09:44:15  &   6  & $59.12469$ &  $0.99873$  &  $-0.00008$\\
29  & 09:44:51  &  09:45:00  &   9  & $58.97522$ &  $0.99620$  &  $-0.00022$\\
30  & 09:45:21  &  09:45:00  &  21  & $58.94966$ &  $0.99577$  &  $-0.00025$\\
31  & 09:45:51  &  09:45:45  &   6  & $59.03442$ &  $0.99720$  &  $-0.00017$\\
32  & 09:46:21  &  09:46:30  &   9  & $58.97185$ &  $0.99615$  &  $-0.00023$\\
33  & 09:46:51  &  09:46:30  &  21  & $58.98279$ &  $0.99633$  &  $-0.00022$\\
34  & 09:47:21  &  09:47:15  &   6  & $58.96922$ &  $0.99610$  &  $-0.00023$\\
35  & 09:47:50  &  09:48:00  &  10  & $58.98497$ &  $0.99637$  &  $-0.00022$\\
36  & 09:48:20  &  09:48:00  &  20  & $59.03450$ &  $0.99720$  &  $-0.00017$\\
37  & 09:48:50  &  09:48:45  &   5  & $59.04133$ &  $0.99732$  &  $-0.00016$\\
38  & 09:49:20  &  09:49:30  &  10  & $58.98776$ &  $0.99641$  &  $-0.00021$\\
39  & 09:49:50  &  09:49:30  &  20  & $59.03244$ &  $0.99717$  &  $-0.00017$\\
\bottomrule
mean: & & & 12.3 & $58.994 \pm 4.5 \times 10^{-2}$ & $0.99652$ & $-0.00021$
\end{tabular}
\end{minipage}}
\end{table}

\subsection{Final Uncertainty}

\vspace{\baselineskip}\noindent
The uncertainty obtained through the standard deviation of the distribution is of course only a part of the total uncertainty, as it merely takes into consideration how the image scale varies from image pair to image pair. To find the total uncertainty we must also account for the uncertainty that is associated with the SDO/HMI calculations, which as previously stated, are dependent upon on their estimate of the solar radius ($R_{\mathrm{ref}}$).

The standard deviation of the distribution ($4.5\times10^{-5}$) corresponds to $\sim 0.077\%$ of the central value, while the uncertainty of the image scale in the SDO/HMI images used as references varies between 0.008\% in the best-case, or 0.07\% in the worst-case, depending on how well their estimates of $R_{\mathrm{ref}}$ matches reality. Best-case in this context is in reference to a radius of \SI{695945}{\kilo\meter}, which they calculated through the Venus transit~\citep{schou13}. Worst case is in reference to a radius of \SI{695508}{\kilo\meter} \citep{1998ApJ...500L.195B} compared to the currently used value of $R_{\mathrm{ref}} = \SI{696000}{\kilo\meter}$. By quadratically adding the standard deviation of the \ac{sst} with the worst-case, we arrive at a final combined uncertainty in the SST image scale of $\lesssim 0.1\%$.

Figure~\ref{image:chi2minimum} shows the well shaped $\chi^2$ minimum for one image pair, which serves to demonstrate that the optimization is unlikely to get stuck in local minima and that there actually is a minimum to optimize the zoom-factor (hence the scale factor) towards within the given uncertainty. The best-fit image scale is shown as the red vertical line. While the old image scale, that was determined during the 2004 Venus transit, is shown as a blue vertical line. The green and dark green lines correspond to the uncertainties associated with the best- and worse case estimates of the solar radius~\citep{schou13}, while the histogram illustrates the approximate normal distribution of the SST image scale throughout the 40 image pairs.

\begin{sidewaysfigure}
	\includegraphics[width=\textwidth]{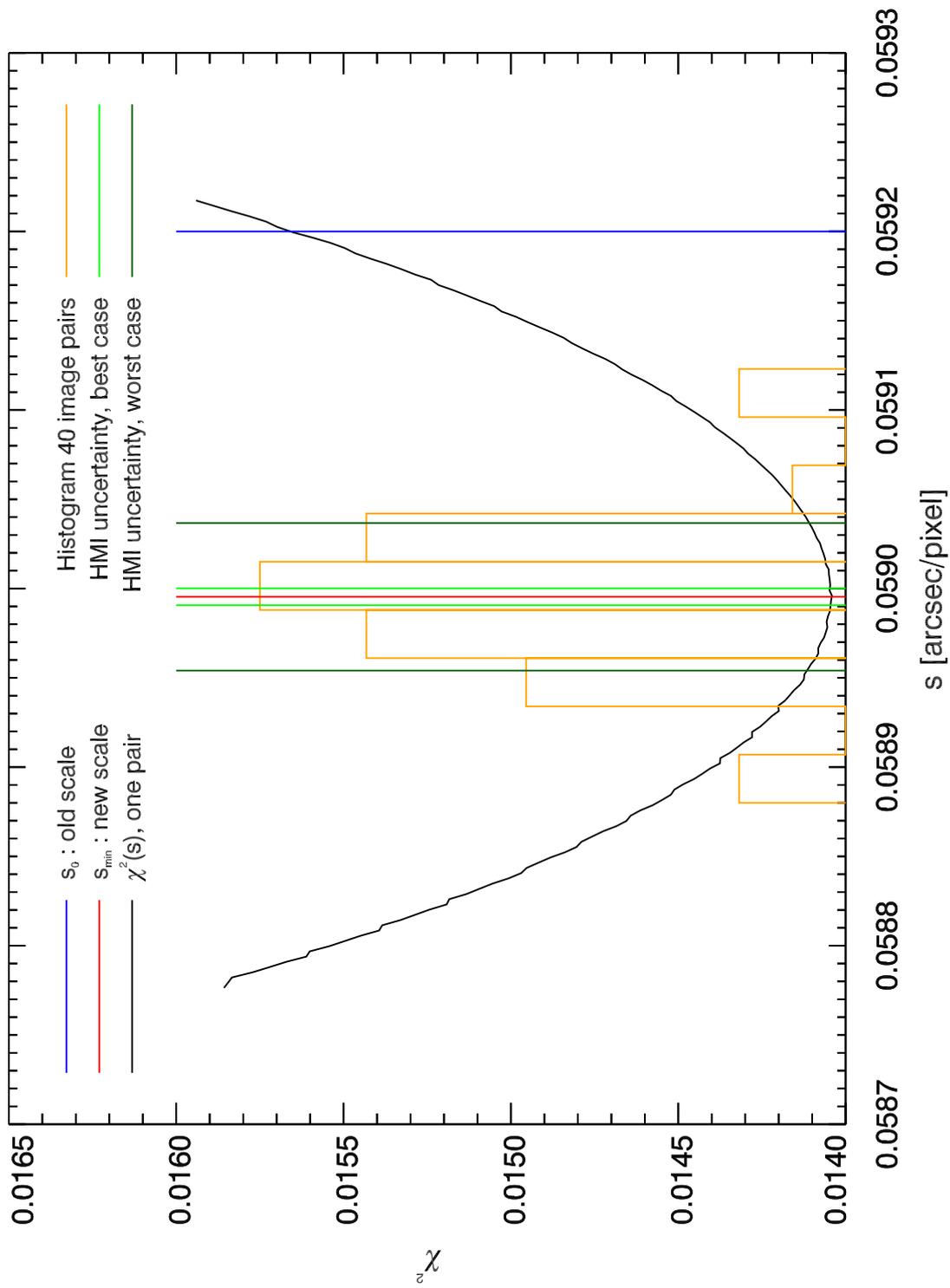}
		\caption[$\chi^2$ minimum]{$\chi^2$ minimum for one image pair including image scales with uncertainties as well as a histogram over the 40 		pairs.}
	\label{image:chi2minimum}
\end{sidewaysfigure}

\newpage
\section{Discussion and Conclusion}

Review: This method is based of the following procedure. Loading relevant SST \& HMI images and then degrading the SST image into a similar resolution as that of the HMI. When these images are of similar resolution, the images may be compared to one another by means of minimizing the least square of the pixel intensity between them. The amount of zoom required in going from the original FOV of the HMI to the same FOV as that of the SST can then be used to calculate the image scale of the SST camera, since the image scale of the HMI is fairly well determined.

\vspace{\baselineskip}\noindent
As it turns out, the largest uncertainty is the one associated with the standard deviation of the distribution of image pairs. This uncertainty 
will of course vary somewhat in future determinations, since no image pair is the other like, though it should remain relatively close to the current deviation of $\sim 0.077\%$ of the central value. However, with a larger sample of images the uncertainty may conceivably go down. This does not mean that the uncertainty associated with the SDO/HMI image scale is negligible in comparison. On the contrary, it is similar in percentage to the central value of the HMI image scale ($\sim 0.505$ arcsec/pixel), namely $\sim 0.07\%$. These adds up to a combined uncertainty of approximately $0.1\%$ of the central value of the SST image scale. The good news is that as the value of solar radius $R_{\mathrm{ref}}$ is further refined, the uncertainty of the \ac{hmi} image scale will go down, and with it the uncertainty in the \ac{sst} image scale.

Using this method, the scale factor of the SST/CRISP camera has been calculated to be $0.05899 \pm 6 \times 10^{-5}$. Which confirms that the method is viable and fulfills the initial goal of having a method to routinely measure the image scale without the need for the telescope to be opened at night. This value will also serve as a reference value in order to calculate a corresponding grid spacing of the pinhole array, which in turn gives the ability to determine the image scale for any future observations in all of the science-cameras.

\vspace{\baselineskip}\noindent
In conclusion: This work has validated this idea as a suitable replacement to the transit method and means we have revised the image scale of the CRISP camera by about $0.35\%$ to the new value of: $0.05899 \pm 6 \times 10^{-5}$. Corresponding to a grid spacing of the pinhole array of $5\farcs15$.

\vspace{\baselineskip}\noindent
The results of this work have also been presented as an e-poster during the 1st SOLARNET - 3rd EAST/ATST meeting in Oslo, Norway~ \citep{noren13poster}. \textit{Please note: Some improvements to the degradation process means that the values have been slightly changed since}.
}

\newpage
\section{Acknowledgments}

I would like to thank the faculty of the Institute for Solar Physics at Stockholm University for providing me with such an interesting topic for my project thesis. I'd especially like to extend my deepest gratitude to Mats Löfdahl, the supervisor of this thesis. Whom have been a guiding hand throughout the course of this project, and whose expertise, eager helpfulness and easygoing demeanor has proven invaluable and made this project both a learning and enjoyable experience. 

I'd also like to thank my friends and family for their encouragement and unwavering moral support. A special thanks also goes out to my friend Carl-Douglas Benno for being kind enough to proof-read this thesis and providing valuable input.

\newpage
\bibliographystyle{plainnat}
\bibliography{mybib}

\begin{thebibliography}{8}
\providecommand{\natexlab}[1]{#1}
\providecommand{\url}[1]{\texttt{#1}}
\expandafter\ifx\csname urlstyle\endcsname\relax
  \providecommand{\doi}[1]{doi: #1}\else
  \providecommand{\doi}{doi: \begingroup \urlstyle{rm}\Url}\fi

\bibitem[{Brown} and {Christensen-Dalsgaard}(1998)]{1998ApJ...500L.195B}
T.~M. {Brown} and J.~{Christensen-Dalsgaard}.
\newblock {Accurate Determination of the Solar Photospheric Radius}.
\newblock \emph{\apjl}, 500:\penalty0 L195, June 1998.
\newblock \doi{10.1086/311416}.

\bibitem[{Garc{\'{\i}}a Mu{\~n}oz} and {Mills}(2012)]{2012A&A...547A..22G}
A.~{Garc{\'{\i}}a Mu{\~n}oz} and F.~P. {Mills}.
\newblock {The June 2012 transit of Venus. Framework for interpretation of
  observations}.
\newblock \emph{\aap}, 547:\penalty0 A22, November 2012.
\newblock \doi{10.1051/0004-6361/201219738}.

\bibitem[{Markwardt}(2009)]{2009ASPC..411..251M}
C.~B. {Markwardt}.
\newblock {Non-linear Least-squares Fitting in IDL with MPFIT}.
\newblock In D.~A. {Bohlender}, D.~{Durand}, and P.~{Dowler}, editors,
  \emph{Astronomical Data Analysis Software and Systems XVIII}, volume 411 of
  \emph{Astronomical Society of the Pacific Conference Series}, page 251,
  September 2009.

\bibitem[Nor{\'e}n and L{\"o}fdahl(2013)]{noren13poster}
Alexander Nor{\'e}n and Mats L{\"o}fdahl.
\newblock Image scale calibration for the swedish 1-m solar telescope.
\newblock In \emph{Synergies Between Ground and Space Based Solar Research, 1st
  SOLARNET -- 3rd EAST/ATST meeting}, 2013.
\newblock Oslo, August.

\bibitem[{Scharmer} et~al.(2003){Scharmer}, {Kiselman}, {L{\"o}fdahl}, and
  {Rouppe van der Voort}]{2003ASPC..307....3S}
G.~B. {Scharmer}, D.~{Kiselman}, M.~G. {L{\"o}fdahl}, and L.~H.~M. {Rouppe van
  der Voort}.
\newblock {First Results from the Swedish 1-m Solar Telescope}.
\newblock In J.~{Trujillo-Bueno} and J.~{Sanchez Almeida}, editors, \emph{Solar
  Polarization}, volume 307 of \emph{Astronomical Society of the Pacific
  Conference Series}, page~3, 2003.

\bibitem[{Schou} et~al.(2012){Schou}, {Scherrer}, {Bush}, {Wachter},
  {Couvidat}, {Rabello-Soares}, {Bogart}, {Hoeksema}, {Liu}, {Duvall}, {Akin},
  {Allard}, {Miles}, {Rairden}, {Shine}, {Tarbell}, {Title}, {Wolfson},
  {Elmore}, {Norton}, and {Tomczyk}]{2012SoPh..275..229S}
J.~{Schou}, P.~H. {Scherrer}, R.~I. {Bush}, R.~{Wachter}, S.~{Couvidat}, M.~C.
  {Rabello-Soares}, R.~S. {Bogart}, J.~T. {Hoeksema}, Y.~{Liu}, T.~L. {Duvall},
  D.~J. {Akin}, B.~A. {Allard}, J.~W. {Miles}, R.~{Rairden}, R.~A. {Shine},
  T.~D. {Tarbell}, A.~M. {Title}, C.~J. {Wolfson}, D.~F. {Elmore}, A.~A.
  {Norton}, and S.~{Tomczyk}.
\newblock {Design and Ground Calibration of the Helioseismic and Magnetic
  Imager (HMI) Instrument on the Solar Dynamics Observatory (SDO)}.
\newblock \emph{\solphys}, 275:\penalty0 229--259, January 2012.
\newblock \doi{10.1007/s11207-011-9842-2}.

\bibitem[Schou(2013)]{schou13}
Jesper Schou.
\newblock Private communication, February 2013.

\bibitem[{Shine} et~al.(1994){Shine}, {Title}, {Tarbell}, {Smith}, {Frank}, and
  {Scharmer}]{1994ApJ...430..413S}
R.~A. {Shine}, A.~M. {Title}, T.~D. {Tarbell}, K.~{Smith}, Z.~A. {Frank}, and
  G.~{Scharmer}.
\newblock {High-resolution observations of the Evershed effect in sunspots}.
\newblock \emph{\apj}, 430:\penalty0 413--424, July 1994.
\newblock \doi{10.1086/174416}.

\end{thebibliography}
\end{document}